# Spatially resolved spectral phase interferometry with isolated attosecond pulse

HIROKI MASHIKO,[1,*] MING-CHANG CHEN,[2] KOJI ASAGA,[1,3] AKIHIRO OSHIMA,[1,4] IKUFUMI KATAYAMA,[4] JUN TAKEDA,[4] TADASHI NISHIKAWA,[3] AND KATSUYA OGURI,[1]

[1]*NTT Basic Research Laboratories, 3-1 Morinosato Wakamiya, Atsugi, Kanagawa 243-0198, Japan*
[2]*Institute of Photonics Technologies, National Tsing Hua University, Hsinchu 30013, Taiwan*
[3]*Department of Electronic Engineering, Tokyo Denki University, 5 Senju-Asahi-cho, Adachi-ku, Tokyo 120-8551, Japan*
[4]*Department of Physics, Yokohama National University, 79-5 Tokiwadai, Hodogaya, Yokohama 240-8501, Japan*
*\*hiroki.mashiko.wv@hco.ntt.co.jp*

**Abstract:** We characterized spatially resolved spectral phase interferometry with an isolated attosecond pulse (IAP). The measured spatial-spectral interferogram and spectral interference fringe visibility show a high degree of IAPs spatial and spectral coherences. In addition, the characterized spectral-delay interferogram shows periodic temporal oscillations over the full IAP continuous spectrum, which indicates high temporal coherence. The IAP coherence over broad continuous spectral region holds potential for realizing the time-resolved IAP phase-based spectroscopy, which will contribute to exploring spatiotemporal dispersive electronic wave dynamics in the future.

## 1. Introduction

High-order harmonics in the extreme ultraviolet (XUV) and soft x-ray regions can have high spatial and temporal coherence [1]. Combining spatial and temporal detection by means of self-referencing interferometry has a variety of applications in, for instance, molecular orbital tomography [2], wavefront reconstruction [3], nanoscale sample imaging [4-6], and electric field characterization of attosecond pulse [7].

To date, the attosecond phase-based spectroscopy for observing electronic wave have been reported by using two phase-locked attosecond pulse trains (APTs) [2, 8-10]. Since the APT has discrete odd-order harmonic spectra [11], an isolated attosecond pulse (IAP) with continuous spectrum is further useful for characterizing the electronic wave. In recent year, an attosecond streaking spectroscopy using phase-locked IAP and near-infrared (NIR) pulse have been used to measure the static relative group delay dispersion between photoelectrons emitted from ground and shake-up states in helium atom, and that was observed over full IAP continuum spectral region [12].

Meanwhile, in traditional time-resolved phase-based spectroscopy in a femtosecond near-infrared (NIR) region, the reference pulse is commonly used, in addition to pump and probe pulses [13-15], which allows direct detection of the transient complex response of samples, both real and imaginary parts. The advantage of this method allows us to select an arbitral temporal gate width between reference- and probe-pulses, and to perform continuous delay scan between the pump-pulse and the pair of reference- and probe-pulses. To implement this concept to attosecond region, the capability of spectral phase interferometry with two IAPs is key issue. However, it is still technically challenge due to, for instance, the wavefront distortion in high-order harmonic generation (HHG) process [16, 17] and the chromatic aberrations in broadband spectrum [18], in addition to the robust interferometer construction in XUV region. In this work, we successfully characterized spatially resolved spectral phase interferometry using IAP with broad continuous spectrum. The measured spatial-spectral and spectral-delay interferograms indicate a high degree of IAPs coherence in space, time, and frequency domains, which shows the potential of realizing time-resolved phase-based spectroscopy with IAPs.

## 2. Experiment

For the HHG, we used the carrier-envelope phase and a beam-pointing-stabilized few-cycle near-infrared (NIR) driving pulse (1.57-eV center photon energy, 6-fs duration, 500-µJ pulse energy, and 3-kHz repetition rate) from a Ti:sapphire laser. The IAP was generated using the double optical gating (DOG) technique [19, 20]. The DOG field is constructed by two quartz plates (270- and 480-µm-thick) and the $\beta$-BaB$_2$O$_4$ (BBO) crystal (140-µm-thick), which has designed temporal gate width of approximately 1.3 fs (half-cycle of NIR pulse). The NIR driving pulse with 9-mm in diameter is focused by a spherical mirror (radius of curvature of 800 mm) into the gas cell (3-mm interaction length) filled with argon gas. The estimated spot size and Rayleigh length are 22 µm and 1.9 mm, respectively. In the optimized phase matching condition for higher flux, the backing gas pressure is 9 mbar, and the gas cell location is approximately 2 mm after the geometrical focus position of NIR pulse, respectively. The estimated target peak intensity is $1.6 \times 10^{15}$ W/cm$^2$.

Figure 1(a) shows a schematic view of the IAP spectral phase interferometry setup. To construct the interferometer, we used a reflection beam-splitting mirror (BSM) in this experiment. Since the use of a transmission beam splitter is challenging due to strong absorption in the XUV region, a reflection BSM [21-24] or a two-HHG-source scheme [2-7] is commonly used to build an XUV interferometer. The advantage of directly splitting a single IAP into two pulses using the reflection BSM is that the relative phase between IAPs can be simply defined. In this experiment, the BSM was made of fused silica without a coating, which is equipped with a piezoelectric transducer (PZT) with 1-nm distance resolution (SmarAct inc.: SLC-1720). It was used in a grazing incidence configuration with the angle of incidence (AOI) of 85 degrees, which reduces the effective distance and improves temporal resolutions. The PZT resolution of 1 nm at the normal direction of the mirror is effectively improved approximately 11 times to 0.09 nm according to the cosine of 85 degrees, corresponding to 0.6-as delay resolution considering the beam roundtrip path length. The configuration also helps mitigate the mechanical vibration issue in the XUV interferometer.

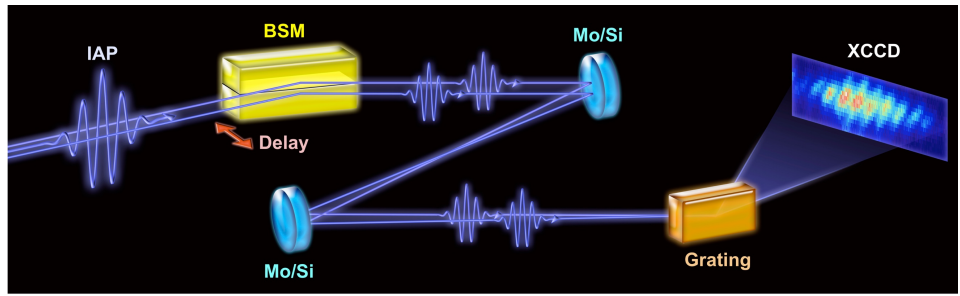

Fig. 1. Schematic view of the spatially resolved phase interferometry setup with IAP. BSM: beam splitting mirror, which has a delay stage function and is equipped with a piezo-electric transducer with 1-nm distance resolution. Mo/Si: Mo/Si multilayer-coated spherical mirrors with curvature radiuses of 500 and 400 mm (10% reflectivity at 25–70 eV). Grating: XUV diffraction grating with 600 lines/mm. XCCD: X-ray CCD camera with 13.5-µm pixel resolution.

The split IAPs were sent to the first molybdenum silicon (Mo/Si) multilayer-coated spherical focusing mirror (radius of curvature of 500 mm; reflectivity of 10% at 25–70-eV photon energy). The effective focal length was 340 mm considering with the beam divergence from the HHG gas cell. The estimated spot size is 7 µm at zero delay between IAPs. In the focus position, the target sample will be installed for further applications. After the focus, IAPs were sent to the second Mo/Si multilayer-coated spherical focusing mirror (radius of curvature: 400 mm), which has the function of image transfer optics. The IAPs were refocused and spatially overlapped on the X-ray CCD (XCCD) camera (Newton SO DO940P, Andor Oxford Instruments Inc.) in the XUV spectrometer. Although the spectral phase interferogram can be observed by the diverged beam in principle, the focused beam has higher power density that allows smaller area detection, and therefore sufficiently reduce the detector noise of XCCD camera. The object and image distances are 218 and 1850 mm, respectively. The magnification is 8.5 times. The size of a single pixel of the XCCD camera is 13.5 μm. The spectrometer resolution is 150 meV at 45.5-eV photon energy, which was calibrated by autoionizing state in neon atom (electronic transition from $2s^22p^6$ to $2s^22p^6(2S_{1/2})3p$) [25]. In this system, the measured delay jitter between IAPs and beam position stabilities were 2.6-as root mean square (rms) and 0.2-μm rms, respectively, which was evaluated from the signal accumulates 3,000 laser shots (for details of the stability evaluation, see appendix A). In addition, the generated IAP has 386-as coherence time, which was characterized by the first-order interferometric autocorrelation using this system (for details of the autocorrelation, see appendix B).

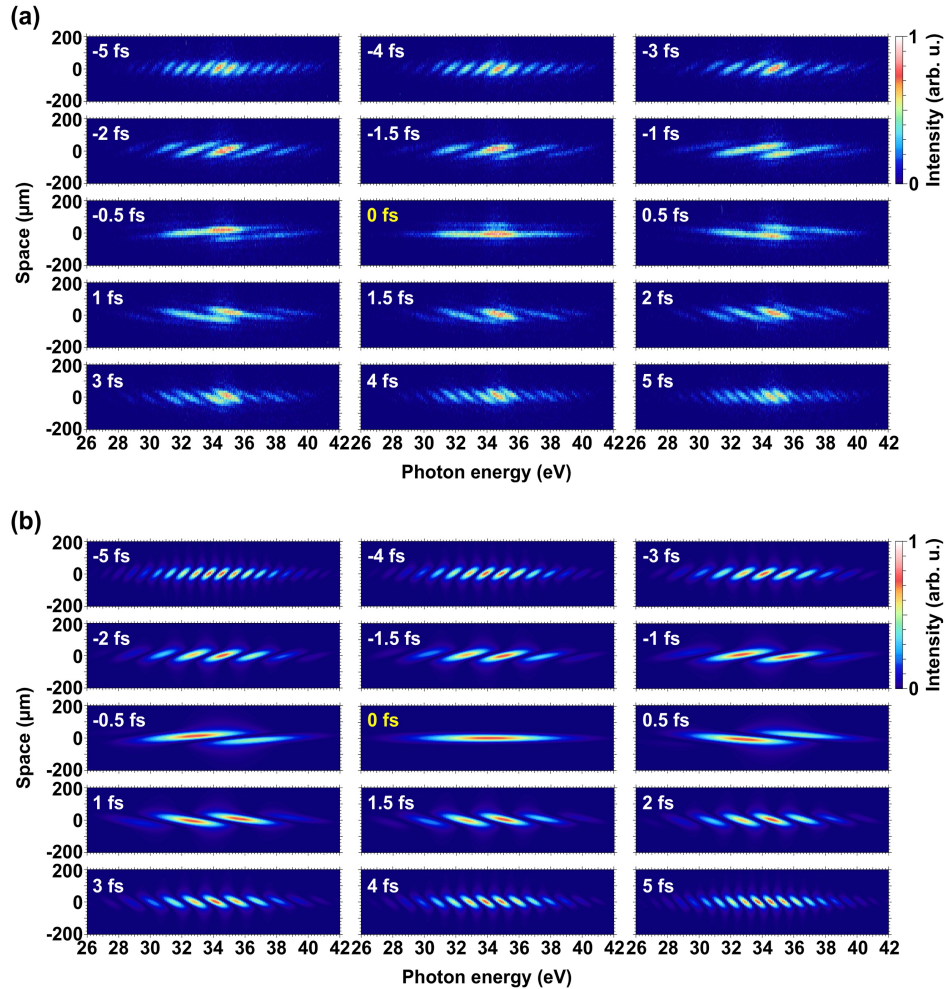

Fig. 2. Spatial-spectral interferograms with spatially split IAPs. (a) Measured and (b) calculated spatial-spectral interferograms over ±5-fs delay regions. The negative delay corresponds to IAP1 (splitted top beam in the BSM) earlier than IAP2 (splitted bottom beam in the BSM). The positive delay is the opposite relation. The signal in (a) accumulates 30,000 laser shots each delay step, which is averaged for 30 measurements.

## 3. Result and discussion

Figure 2(a) and (b) show the measured and calculated spatial-spectral interferograms over ±5-fs delay regions, respectively. Here, we refer to the splitted top and bottom beams by the BSM as IAP1 and IAP2, respectively. The negative delay corresponds to IAP1 (splitted top beam) earlier than it does to IAP2 (splitted bottom beam). The positive delay is the opposite relation. The delay-dependent spatial-spectral interferograms were observed in the wide continuum spectrum of the IAP, as shown in Fig.2(a), which are agrees well with numerically calculated results based on the diffraction theory [26] (for details of the calculation, see appendix C). The Fourier-transform-limited pulse estimated from the spectrum at zero delay between IAPs has 257-as duration. In the case of the negative delay, the spatial-spectral interferogram presents upward to the right, while the interferogram shows a straight line at the zero delay. In the case of the positive delay, it presents downward to the right. Asymmetrical spatial-spectral interferograms bounded on the zero delay are caused by IAP spatial phases (wavefronts) tilted in opposite directions [27]. The spatial phase effect is precisely shown in the measured spatial-spectral interferograms. The time-dependent spatial fringe is an important interferometric component, which can extract the spatial phase information from Fourier transform. It can relax the constraint of the spectral resolution of the XUV spectrometer [4, 7] and be used for sample imaging [6].

The visibility of the spectral interference fringe is important for the investigation of the IAP coherence. The interference fringe visibility $V(\omega)$ is given by [28]

$$V(\omega) = \frac{I_0(\omega) - I(\omega)}{I_0(\omega) + I(\omega)}, \quad (1)$$

where $I_0(\omega)$ is the lineout spectral interferogram with zero delay at center 0-μm space in the spatial-spectral interferogram. The $I(\omega)$ corresponds to lineout spectral interferogram with delay between IAPs. Figure 3 shows spectral interference fringe visibilities for delays from -1 to -6 fs over photon energy regions of 32–38 eV. The energy gap $\Delta E$ of periodic modulation in the spectral interference fringe is expressed as $\Delta E = 2\pi\hbar/\tau$, where $\hbar$ and $\tau$ are Dirac's constant and the delay between IAPs, respectively. The energy gap corresponds to 4.1 eV at -1 fs, 2.1 eV at -2 fs, 1.4 eV at -3 fs, 1 eV at -4 fs, 0.8 eV at -5 fs, and 0.7 eV at -6 fs, which are represented well in Fig. 3. The maximum visibility is approximately 0.6 in our optimized phase matching condition of HHG. The visibility could be reduced by the imperfect phase-matching condition, which produces the spatiotemporal phase distortion [16-18]. In addition, the limitation of the spectral and spatial resolutions in the XUV spectrometer could also cause the visibility reduction. Nevertheless, the interference fringe visibility reaches a certain level of quality in the XUV region, which can apply for the Fourier spectroscopy.

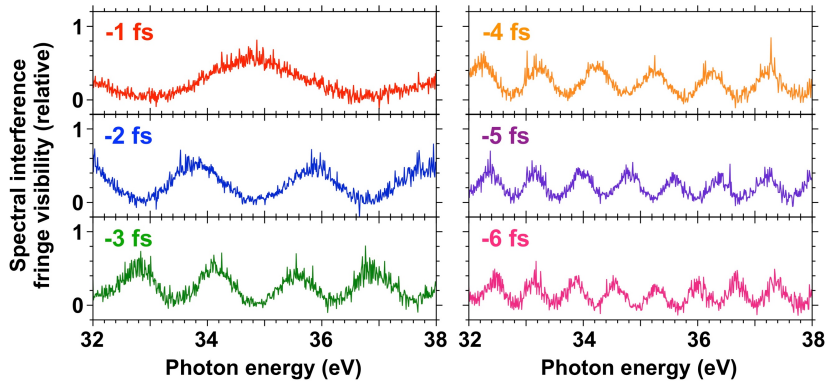

Fig. 3. Spectral interference fringe visibilities with IAP. Visibilities correspond to delays from -1 to -6 fs over photon energy regions of 32–38 eV. The maximum visibility is approximately 0.6. The energy gaps $\Delta E$ of periodic modulation in the spectral interference fringe are 4.1 eV at -1 fs, 2.1 eV at -2 fs, 1.4 eV at -3 fs, 1 eV at -4 fs, 0.8 eV at -5 fs, and 0.7 eV at -6 fs, respectively.

Figure 4(a) shows measured spectral-delay interferograms with the IAP. The delay step corresponds to 12 as. The spectral-delay interferograms can be obtained from the lineout spectrum for the spatial-spectral interferogram at each delay in Fig. 2. The spectral-delay interferogram shows periodic temporal oscillations of 103–159 as, which indicates high temporal coherence over the full IAP bandwidth of 26–40 eV. Figure 4(b) shows spectral-spectral interferograms after Fourier transformation for the delay axis in (a). Alternating-current (AC) components appear in ±26–40-eV regions, in addition to the direct-current (DC) component around zero photon energy. Since the AC component has the continuum spectrum corresponding to the IAP, the temporal oscillations in (a) is projected well onto the spectral-spectral interferogram in (b). In this case, a Fourier window can be applied for the AC components. This analytical process is important for removing a DC component including the detector shot noise, which helps for the low photon flux issue of attosecond light sources [29] comparing with a synchrotron radiation source [30] and an x-ray free electron laser [31]. In addition, since the AC component has information of the relative intensity profile and phase difference between IAPs, corresponding to real and imaginary parts, the spectral phase interferogram will be a powerful tool for nuclear and electron dynamics detection through time-resolved phase-based spectroscopy with IAPs in the future.

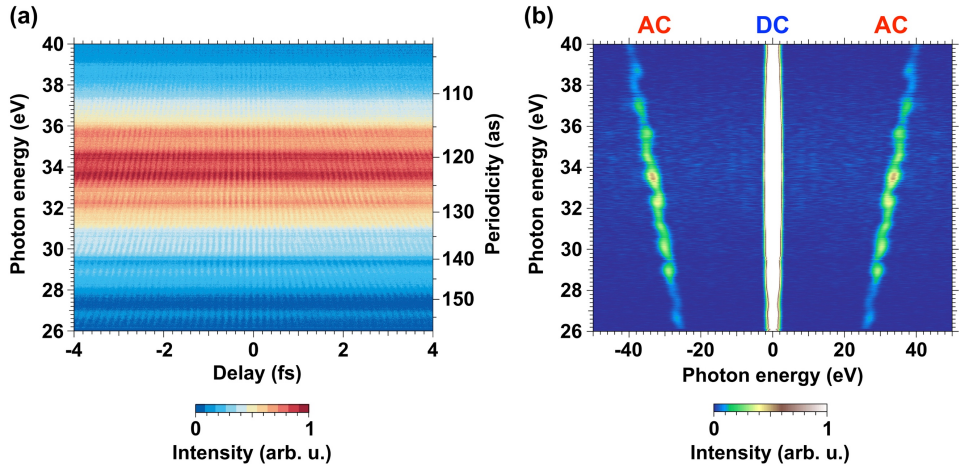

Fig. 4. Spectral-delay interferograms with IAP. (a) Measured spectral-delay interferograms. The delay step is 12 as. The signal accumulates 6,000 laser shots each delay step, which is averaged for 17 measurements. The periodic temporal oscillations correspond to 103–159 as over the bandwidth of 26–40 eV. (b) Spectral-spectral interferograms after Fourier transformation for delay axis in (a). AC components appear in ±26–40-eV regions, in addition to the DC component around zero photon energy.

## 4. Conclusion

We characterized spatially resolved spectral phase interferometry with the IAP. The spatial-spectral interferogram with spatially split IAPs shows asymmetrical interferogram patterns with respect to the zero delay. According to space-time phase coupling, the spatial phase (wavefront) effect was observed in the measured spatial-spectral interferograms, which indicates high IAP spatial and spectral coherences. The visibility of the spectral interference fringe reaches 0.6 in our optimized condition. In addition, the measured spectral-delay interferogram shows periodic temporal oscillations over the full IAP bandwidth, which indicates high degree of temporal coherence. The IAP coherence over broadband continuous spectral region holds potential for realizing the time-resolved phase-based spectroscopy using IAPs, which will be contributed in exploring spatiotemporal dispersive electronic wave dynamics in a variety of atomic, molecular, and solid systems in the future.

## Appendix A: Delay and beam position stabilities

This interferometric experiment using IAPs in the XUV region requires very high temporal and spatial stabilities. We evaluated the delay and the beam position stabilities from the spatial-spectral interferogram. Figure 5(a) shows the measured interferogram for delay of -3.2 fs. In this evaluation, we used IAPs with higher photon energy (36–50 eV) than the ones we used previously (Fig. 2, 3, and 4) because the interferogram with a shorter wavelength has higher sensitivity. The upper image in (b) shows lineout spectra from the integrated spatial area of ±13.5 μm in (a). The signal accumulates 3,000 laser shots every spectrum, which corresponds to 1-s exposure time in the XCCD. The value is similar to the exposure time in the previous measurements in Fig. 2, 3, and 4. The bottom graph shows delay and relative phase jitters extracted from the upper image through the Fourier analysis. The rms values of the delay and relative phase jitter measured over 30 min. are 2.6-as and 172-mrad, respectively. These values are significantly smaller than the IAP optical cycle of approximately 120-as periodicity in the XUV region.

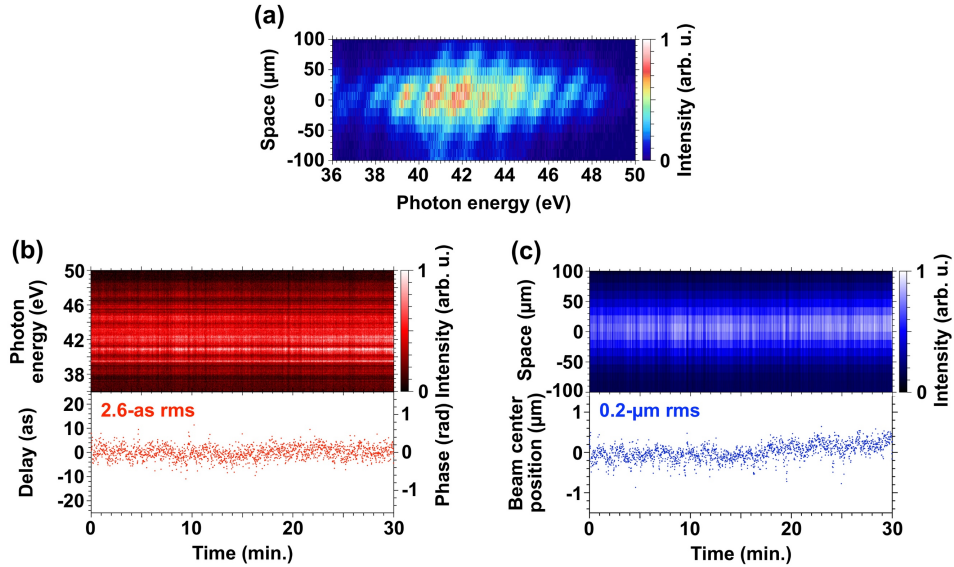

Fig. 5. Delay and beam position stabilities with IAP. (a) Measured spatial-spectral interferogram for a delay of -3.2 fs. (b) Top image: lineout spectra integrated spatial area of ±13.5 μm in (a). The signal accumulates 3,000 laser shots every spectrum. Bottom graph: delay and relative phase jitters between IAPs extracted from the upper figure through the Fourier analysis. The root mean squire (rms) values of the delay and relative phase jitter measured over 30 min. are 2.6-as and 172-mrad, respectively. (c) Top image: beam profiles integrated photon energy regions over 36–50 eV in (a). The signal accumulates 3,000 laser shots in every profile, which is synchronized for taking data of (b). Bottom graph: Gaussian fitted peak value for beam profiles. The beam center position stability measured over 30 min. corresponds to 0.2-μm rms.

Furthermore, to evaluate the stability in the spatial domain, the beam position with the interferogram was monitored. The upper part of (c) shows lineout beam profiles, which are integrated photon energy regions over 36–50 eV in (a). The signal accumulates 3,000 laser shots every beam profile, which is synchronized for taking data of (b). The bottom graph shows Gaussian fitted peak value for the beam profile. The beam center position stability measured over 30 min. corresponds to 0.2-µm rms. In this experiment, the distance separation of the spatial interference fringe was approximately 60 µm at 42-eV center photon energy. If the stability is larger than the distance separation, the interferogram will be smeared. Consequently, the high spatial and temporal stabilities allow us to resolve the spectral phase interferogram.

**Appendix B: First-order interferometric autocorrelation**

Figure 6(a) shows the measured IAP spectrum at zero delay between IAPs. The Fourier-transform-limited pulse estimated from the spectrum has 257-as duration. Figure 6(b) shows the measured first-order interferometric autocorrelation trace. The inset shows the trace enlarged over the ±5-fs region, which indicates the pulse isolation. The blue dashed line in (b) corresponds to the reconstructed first-order interferometric autocorrelation trace from (a) through the inverse Fourier transformation, which agrees well with the measured trace (red filled circles and solid line) in (b). The estimated coherence time from the measured trace is 386 as.

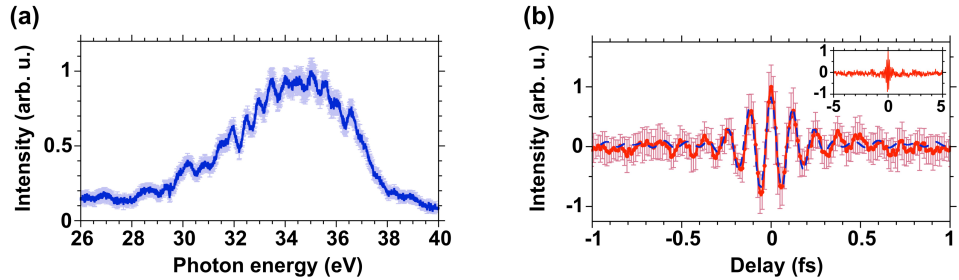

Fig. 6. IAP spectrum and first-order interferometric autocorrelation. (a) Blue solid line shows the measured IAP spectrum at zero delay between IAPs. The Fourier-transform-limited pulse estimated from the spectrum has 257-as duration. (b) Measured first-order interferometric autocorrelation trace (red filled circles and solid line), which was observed without a diffraction grating. The delay step is 12 as. The inset shows the trace enlarged over the ±5-fs region. The blue dashed line is the spectrum reconstructed from (a) through inverse Fourier transformation. The estimated coherence time from the measured trace is 386 as. Error bars in (a) and (b) represent the root mean square (rms) over 15 measurements.

**Appendix C: Calculation of spatial-spectral interferogram**

The spatially split Gaussian beam has tilted spatial phases (wavefronts) at the focal point [27], which affects the spatial-spectral interferogram. Figure 7(a) shows the calculated beam intensity profile with spatially split IAPs, where we refer to the top beam and bottom beam as IAP1 and IAP2, respectively. The ($x_1$, $y_1$) coordinate corresponds to the space on the first Mo/Si mirror after the beam splitting mirror (BSM) in the experiment. In this calculation, the input IAP beam radius (before beam splitting) and the focal length are 563 μm and of 340 mm, respectively. These values are estimated from the experimental condition. Figure 7(b) shows images of focused beam intensity profiles (left) and spatial phases (middle), which are numerically calculated based on a diffraction theory [26]. The ($x_2$, $y_2$) coordinate corresponds to the space on the XCCD camera in the experiment. Note that these images already consider the image transfer of 8.5 times magnification in this experiment, i.e., they fit the image on the XCCD camera. The upper and lower beams correspond to IAP1 (top beam) and IAP2 (bottom beam), respectively. The right graphs show lineout intensity profiles (red solid line) and phases (blue dashed line) at the space $x_2$=0 μm in the left and middle images. Although IAPs have the same beam intensity profiles at the focal point, the spatial phases tilt in opposite directions from each other. This is an important point for constructing the spatial-spectral interferogram with spatially split IAPs. The synthesized electric field at the focal point is expressed as [27]

$$\tilde{E}(x_2,y_2,t) = f(t)|\tilde{U}_1(x_2,y_2)|e^{i\varphi_1(x_2,y_2)} + f(t-\tau)|\tilde{U}_2(x_2,y_2)|e^{i\varphi_2(x_2,y_2)}e^{i\Delta\phi}, \quad (2)$$

where the first term on the right-hand side denotes IAP1 and the second term is IAP2 with the delay $\tau$. $f(t)$ is the temporal profile of the pulse envelope. The focused beam intensity profile and the spatial phase are expressed as $\tilde{U}(x_2,y_2)$ and $\varphi(x_2,y_2)$, respectively. The synthesized electric field $\tilde{E}(x_2,y_2,t)$ is modified by the relative temporal phase $\Delta\phi$ (=$\omega\tau$), i.e., the delay $\tau$ between IAPs and the angular frequency $\omega$. The delay-dependent spatial-spectral interferogram in Fig. 2(b) is obtained from the synthesized electric field $\tilde{E}(x_2,y_2,t)$ with the space-time phase coupling through the Fourier transformation for time $t$.

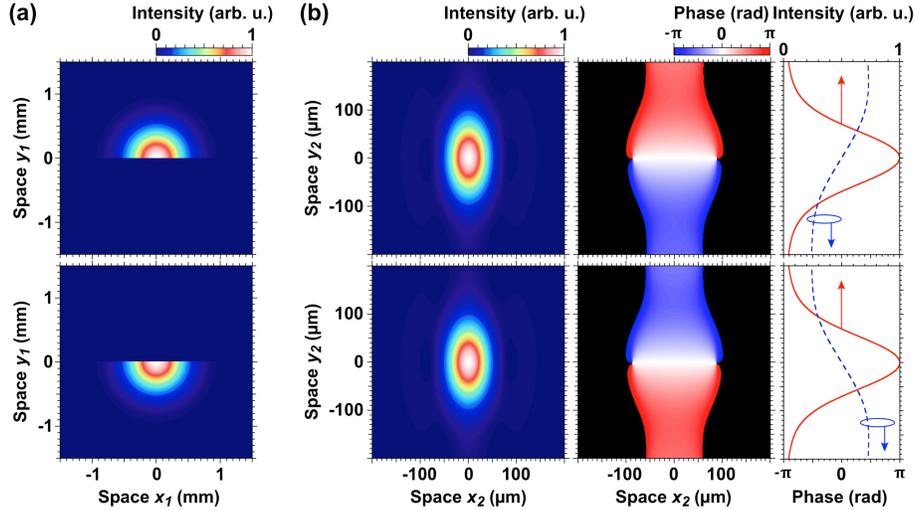

Fig. 7. Calculated spatial beam intensity profile and phase with spatially splitted IAPs. (a) Beam intensity profile of input IAPs (IAP1 for top beam and IAP2 for bottom beam). The ($x_1$, $y_1$) coordinate corresponds to the space on the first Mo/Si mirror after the BSM in the experiment. The input IAP beam radius (before beam splitting) is 563 μm. (b) Images of focused beam intensity profiles (left) and spatial phases (middle), which are numerically calculated based on a diffraction theory [26]. The ($x_2$, $y_2$) coordinate corresponds to the space on the XCCD camera in the experiment. The image transfer of 8.5 times magnification is already considered for images of focused beam intensity profiles and spatial phases. The right graphs show lineout beam intensity profiles (red solid line) and special phases (blue dashed line) at the space $x_2$=0 μm in left and middle images. The upper and lower images and graphs correspond to IAP1 (top beam) and IAP2 (bottom beam), respectively.

**Funding**

This work was supported by JSPS KAKENHI Grant No. 16H05987, 16H02120, and 19H02637 and by Taiwan MOST Grant No. 109-2636-M-007-008.

**Disclosures**

The authors declare no conflicts of interest.